\documentstyle[12pt]{article}

\newcommand{\beq}{\begin{equation}}
\newcommand{\eeq}{\end{equation}}
\newcommand{\bea}{\begin{eqnarray}}
\newcommand{\eea}{\end{eqnarray}}
\newcommand{\gf}{G_{\mbox{{\scriptsize F}}}}
\newcommand{\heff}{H_{\mbox{{\scriptsize eff}}}}
\newcommand{\bsbar}{\overline{B_s}}
\newcommand{\lab}{\label}
\newcommand{\order}{{\cal O}}
\newcommand{\aftilde}{A_{\tilde f}^\ast}
\newcommand{\aftildebar}{\overline{A}_{\tilde f}^\ast}
\newcommand{\af}{A_f}
\newcommand{\afbar}{\overline{A}_f}
\newcommand{\etacp}{\eta_{\mbox{{\tiny CP}}}^f}
\newcommand{\etacptilde}{\eta_{\mbox{{\tiny CP}}}^{\tilde f}}
\newcommand{\xif}{\xi_f}
\newcommand{\xiftilde}{\xi_{\tilde f}}
\newcommand{\deltap}{\delta_{P'}^f}
\newcommand{\deltaptilde}{\delta_{P'}^{\tilde f}}
\newcommand{\rf}{r_f}
\newcommand{\rftilde}{r_{\tilde f}^\ast}
\newcommand{\pf}{P_f'}
\newcommand{\pftilde}{P_{\tilde f}'}
\newcommand{\nff}{N_{f,\tilde f}}
\newcommand{\xx}{(X_1\,X_2)_f}
\newcommand{\xxtilde}{(X_1\,X_2)_{\tilde f}}

\begin{document}

\setcounter{page}{0}
\thispagestyle{empty}

\rightline{TTP96-07}
\rightline{FERMILAB-PUB-96/079-T}
\rightline{hep-ph/9605220}
\rightline{May 1996}
\bigskip
\vspace{0.5cm}
\begin{center}
{\Large {\bf CP violation and CKM phases from}}
\end{center}
\begin{center}{\Large{\bf angular distributions for $B_s$ decays}}
\end{center}
\begin{center}{\Large{\bf into admixtures of CP eigenstates}}\\
\end{center}
\vspace{0.2cm}
\smallskip
\begin{center}
{\large{\sc Robert Fleischer}}\footnote{Internet: {\tt 
rf@ttpux1.physik.uni-karlsruhe.de}}\\
{\sl Institut f\"{u}r Theoretische Teilchenphysik}\\
{\sl Universit\"{a}t Karlsruhe}\\
{\sl D--76128 Karlsruhe, Germany}\\ 
\vspace{0.7cm}
{\large{\sc Isard Dunietz}}\footnote{Internet: {\tt 
dunietz@fnth15.fnal.gov}}\\
{\sl Theoretical Physics Division}\\
{\sl Fermi National Accelerator Laboratory}\\
{\sl Batavia, IL 60510, USA}\\ 
\vspace{0.8cm}
{\large{\bf Abstract}}\\
\end{center}
\vspace{0.1cm}
We investigate the time-evolutions of angular distributions for 
$B_s$ decays into final states that are admixtures of CP-even 
and CP-odd configurations. A sizable lifetime difference between 
the $B_s$ mass eigenstates allows a probe of CP violation in 
time-dependent untagged angular distributions. Interference effects 
between different final state configurations of $B_s\rightarrow D^{*+}_s\,
D^{*-}_s$, $J/\psi\,\phi$ determine the Wolfenstein parameter $\eta$ 
from {\it untagged} data samples, or -- if one uses $|V_{ub}|/|V_{cb}|$ 
as an additional input -- the notoriously difficult to measure CKM 
angle $\gamma$. Another determination of $\gamma$ is possible by 
using isospin symmetry of strong interactions to relate {\it untagged} 
data samples of $B_s\to K^{\ast+}\,K^{\ast-}$ and $B_s\to K^{\ast0}\,
\overline{K^{\ast0}}$. We note that the {\it untagged} angular 
distribution for $B_s\to\rho^0\phi$ provides interesting information 
about electroweak penguins.

\newpage

\section{Introduction}\lab{intro}

Within the Standard Model \cite{sm} one expects \cite{smexpt} 
a large mass difference $\Delta m\equiv m_H-m_L>0$ 
between the physical mixing eigenstates $B_s^H$ (``heavy'') 
and $B_s^L$ (``light'') of the neutral $B_s$ meson system
leading to very rapid $\Delta mt$--oscillations in data samples
of tagged $B_s$ decays. In order to measure these oscillations, an
excellent vertex resolution system is required which is a formidable
experimental task. However, in a recent paper \cite{dunietz} it has
been shown that it may not be necessary to trace these rapid 
$\Delta mt$--oscillations in order to obtain insights into the fundamental
mechanism of CP violation. The point is that the time-evolution of
{\it untagged} non-leptonic $B_s$ decays, where one does not 
distinguish between initially present $B_s$ and $\bsbar$ mesons, 
depends only on combinations of the two exponents $\exp(-\Gamma_Lt)$
and $\exp(-\Gamma_H t)$ and not on the rapid oscillatory $\Delta mt$--terms. 
Since the width difference $\Delta\Gamma\equiv\Gamma_H-\Gamma_L$
of the $B_s$-system is predicted to be of the order $20\%$ 
of the average $B_s$ width \cite{deltagamma}, interesting
CP-violating effects may show up in untagged rates~\cite{dunietz}.

In the present paper we restrict ourselves to quasi two body 
modes $B_s\to X_1\,X_2$ into final states that are admixtures 
of CP-even and CP-odd configurations. The different case where 
the final states are not admixtures of CP eigenstates but can 
be classified instead by their parity eigenvalues is discussed 
in \cite{df2}, where we present an analysis of angular correlations
for $B_s$ decays governed by $\bar b\to\bar c u\bar s$ quark-level 
transitions. If both $X_1$ and $X_2$ carry spin and continue
to decay through CP-conserving interactions, valuable information can
be obtained from the angular distributions of their decay products. 
Examples for such transitions are
$B_s\to D_s^{\ast +}(\to D_s^+\gamma)\,D_s^{\ast-}(\to D_s^-\gamma)$ 
and $B_s\to J/\psi(\to l^+l^-)\,\phi(\to K^+K^-)$ which allow a
determination of the Wolfenstein parameter $\eta$ 
\cite{wolf} from the time-dependences of their {\it untagged} angular 
distributions as we will demonstrate in a later part of this paper. 
Of course, the formalism developed here applies also to final states 
where the $D^{*\pm}_s$ mesons are substituted by higher resonances, 
such as $B_s\rightarrow D_{s1}(2536)^+ D_{s1} (2536)^-.$  For many 
detector configurations, such higher resonances may be preferable over 
$D^{*\pm}_s$, because of their significant branching fractions into all 
charged final states and because of additional mass-constraints of their 
daughter resonances.

If we use the CKM factor 
\beq\lab{Rb}
R_b\equiv\frac{1}{\lambda}\frac{|V_{ub}|}{|V_{cb}|}
\eeq
with $\lambda=\sin\theta_{\mbox{{\scriptsize C}}}=0.22$ as an additional
input, which is constrained by present experimental data to lie 
within the range $R_b=0.36\pm0.08$ \cite{cleo,blo,al}, $\eta$ fixes the
angle $\gamma$ in the usual ``non-squashed'' unitarity triangle 
\cite{ut} of the CKM matrix \cite{km} through
\beq\lab{gamma}
\sin\gamma=\frac{\eta}{R_b}.
\eeq
Using the isospin symmetry of strong interactions to relate the
$\bar b\to\bar s$ QCD penguin contributions to $B_s\to K^{\ast+}(\to\pi K)\,
K^{\ast-}(\to\pi \overline{K})$ and $B_s\to K^{\ast0}(\to\pi K)\,
\overline{K^{\ast0}}(\to\pi\overline{K})$, another determination 
of $\gamma$ is possible by measuring the corresponding {\it untagged}
angular distributions. This approach is another highlight of our paper.
The formulae describing $B_s\to K^{\ast+} K^{\ast-}$ apply also to 
$B_s\to\rho^0\phi$ if we make an appropriate replacement of variables
providing a fertile ground for obtaining information about
the physics of electroweak penguins. 

This paper is organized as follows: In Section~\ref{evol} we calculate
the time-dependences of the observables of the angular distributions
for $B_s$ decays into final state configurations that are admixtures
of different CP eigenstates. The general formulae derived in 
Section~\ref{evol} simplify considerably if 
the unmixed $B_s\to X_1\,X_2$ amplitude is 
dominated by a single CKM amplitude. This important special case is
the subject of Section~\ref{dom} and applies to an excellent 
accuracy to the decays $B_s\to D_s^{\ast +}\,D_s^{\ast-}$ and 
$B_s\to J/\psi\,\phi$ which are analyzed in Section~\ref{gold}. There
we demonstrate that {\it untagged} data samples of these modes
allow a determination of the Wolfenstein parameter $\eta$, which
fixes the CKM angle $\gamma$ if $R_b$ is known. In Section~\ref{gam}
we present another method to determine $\gamma$ from 
{\it untagged} $B_s\to K^{\ast+}K^{\ast-}$ and $B_s\to K^{\ast0}
\overline{K^{\ast0}}$ decays. The formulae derived there are 
also useful to obtain information about electroweak penguins from
{\it untagged} $B_s\to\rho^0\phi$ events. Finally in Section~\ref{sum}
the main results of our paper are summarized. 

\section{Calculation of the time-evolutions}\lab{evol}

A characteristic feature of the angular distributions for the decays
$B_s\to X_1\,X_2$ specified above is that they depend in general on
real or imaginary parts of the following bilinear combinations of 
decay amplitudes:
\beq\lab{e1}
\aftilde(t)\,\af(t).
\eeq
Here we have introduced the notation 
\beq\lab{e2}
\begin{array}{rcl}
\af(t)&\equiv&A(B_s(t)\to\xx)=\langle\xx|\heff|B_s(t)\rangle\nonumber\\
A_{\tilde f}(t)&\equiv&A(B_s(t)\to\xxtilde)=\langle\xxtilde|\heff|
B_s(t)\rangle
\end{array}
\eeq
for the transition amplitudes of initially, i.e.\ at $t=0$, present $B_s$
mesons decaying into the final state configurations $f$ and $\tilde f$ 
of $X_1\,X_2$ that are both CP eigenstates satisfying 
\beq\lab{e3}
\begin{array}{rcl}
({\cal CP})|\xx\rangle&=&\etacp|\xx\rangle\\
({\cal CP})|\xxtilde\rangle&=&\etacptilde|\xxtilde\rangle
\end{array}
\eeq
with $\etacp,\etacptilde\in\{-1,+1\}$. Here $f$ and $\tilde f$ are 
lables that define the relative polarizations of the two hadrons $X_1$
and $X_2$. The tilde is useful for discussing
the case where different configurations of $X_1\,X_2$ with the 
{\it same} CP eigenvalue are present. To make this point more 
transparent, consider the mode $B_s\to J/\psi\,\phi$
which has been analyzed in terms of the linear polarization amplitudes
\cite{rosner} $A_0(t)$, $A_\parallel(t)$ and $A_\perp(t)$ in \cite{ddlr}.
Whereas $A_\perp(t)$ describes a CP-odd final state configuration, both
$A_0(t)$ and $A_\parallel(t)$ correspond to CP-eigenvalue $+1$, i.e.\
to $\af(t)$ and $A_{\tilde f}(t)$ in our notation (\ref{e2}) 
with $\etacptilde=\etacp=+1$.

The amplitudes describing decays of initially present $\bsbar$ mesons are 
given by 
\beq\lab{e2b}
\begin{array}{rcl}
\afbar(t)&\equiv&A(\bsbar(t)\to\xx)=\langle\xx|\heff|\bsbar(t)
\rangle\nonumber\\
\overline{A}_{\tilde f}(t)&\equiv&A(\bsbar(t)\to\xxtilde)=
\langle\xxtilde|\heff|\bsbar(t)\rangle.
\end{array}
\eeq
Both in these expressions and in (\ref{e2}) the operator
\beq\lab{ham}
\heff=\heff(\Delta B=-1)+\heff(\Delta B=+1)
\eeq
denotes an appropriate low energy effective Hamiltonian with 
\beq
\heff(\Delta B=+1)=\heff(\Delta B=-1)^\dagger
\eeq
and
\beq\lab{e2c}
\heff(\Delta B=-1)=\frac{\gf}{\sqrt{2}}\sum_{j=u,c}v^{(r)}_j
{\cal Q}^j\equiv\frac{\gf}{\sqrt{2}}\sum_{j=u,c}v_j^{(r)}
\left\{\sum_{k=1}^2Q_k^j
C_k(\mu)+\sum_{k=3}^{10}Q_kC_k(\mu)\right\},
\eeq
where $v_j^{(r)}
\equiv V_{jr}^\ast V_{jb}$ is a CKM factor that is different for $b\to d$
and $b\to s$ transitions corresponding to $r=d$ and $r=s$, respectively.
The four-quark operators $Q_k$ can be divided into current-current 
operators $(k\in\{1,2\})$, QCD penguin operators $(k\in\{3,\ldots,6\})$
and electroweak penguin operators $(k\in\{7,\ldots,10\})$, with
index $r$ implicit. Note that these operators 
create $s$ and $d$ quarks for $r=s$ and $r=d$, respectively. The Wilson
coefficients $C_k(\mu)$ of these operators, where $\mu=\order(m_b)$
is a renormalization scale, can be calculated in renormalization group
improved perturbation theory. The reader is referred to a nice recent review
\cite{burasnlo} for the details of such calculations. There numerical 
results for the relevant Wilson coefficients are summarized and the 
four-quark operators $Q_k$ are given explicitly.

\vspace{0.3cm}

Applying the well-known formalism describing $B_s-\bsbar$ mixing 
\cite{dunietz,dr}, a straightforward calculation yields the
following expression for the time-dependence of the bilinear 
combination of decay amplitudes given in (\ref{e1}):
\bea
\lefteqn{\aftilde(t)\,\af(t)=\langle\xxtilde|\heff|B_s\rangle^\ast
\langle\xx|\heff|B_s\rangle}\label{e5}\\
&&\times\left[|g_+(t)|^2+\etacptilde\,\xiftilde^\ast\, g_+(t)\,g_-^\ast(t)
+\etacp\,\xif\, g_+^\ast(t)\,g_-(t)+\etacptilde\,\etacp\,\xiftilde^\ast\,
\xif\,|g_-(t)|^2\right],\nonumber
\eea
where
\bea
|g_\pm(t)|^2&=&\frac{1}{4}\left[e^{-\Gamma_Lt}+e^{-\Gamma_Ht}\pm
2e^{-\Gamma t}\cos(\Delta mt)\right]\lab{e6a}\\
g_+(t)g_-^\ast(t)&=&\frac{1}{4}\left[e^{-\Gamma_Lt}-e^{-\Gamma_Ht}-
2ie^{-\Gamma t}\sin(\Delta mt)\right]\lab{e6b}
\eea
with $\Gamma\equiv(\Gamma_L+\Gamma_H)/2$. The observables $\xif$ 
and $\xiftilde$, which contain essentially all the information needed to 
evaluate the time dependence of (\ref{e5}), are related 
to hadronic matrix elements of the combinations ${\cal Q}^j$ of 
four-quark operators and Wilson coefficients appearing 
in the low energy effective Hamiltonian 
(\ref{e2c}) through
\beq\lab{e7}
\xif=e^{-i\phi_{\mbox{{\tiny M}}}^{(s)}}
\frac{\sum\limits_{j=u,c}v_j^{(r)}\langle 
\xx|{\cal Q}^j|\bsbar\rangle}{\sum\limits_{j=u,c}v_j^{(r)\ast}\langle 
\xx|{\cal Q}^j|\bsbar\rangle},
\eeq
where $\phi_{\mbox{{\scriptsize M}}}^{(s)}\equiv2\,\mbox{arg}(V_{ts}^\ast 
V_{tb})$ is the $B_s-\bsbar$ mixing phase. In order to evaluate 
$\xiftilde$, we have simply to replace $f$ in (\ref{e7}) by $\tilde f$. 
Note that we have neglected the extremely small
CP-violating effects in the $B_s-\bsbar$ oscillations in order to derive
(\ref{e5})-(\ref{e7}) \cite{dunietz}. We shall come back to (\ref{e7}) 
in a moment. 
Let us consider the CP-conjugate processes first. The expression 
corresponding to (\ref{e5}) for initially present $\bsbar$ mesons is 
very similar to that equation and can be written as  
\bea
\lefteqn{\aftildebar(t)\,\afbar(t)=\langle\xxtilde|\heff|B_s\rangle^\ast
\langle\xx|\heff|B_s\rangle}\label{e8}\\
&&\times\left[|g_-(t)|^2+\etacptilde\,\xiftilde^\ast\, g_+^\ast(t)\,g_-(t)
+\etacp\,\xif\, g_+(t)\,g_-^\ast(t)+\etacptilde\,\etacp\,\xiftilde^\ast\,
\xif\,|g_+(t)|^2\right].\nonumber
\eea

\vspace{0.3cm}

In the general case the {\it tagged} angular distribution 
for a given decay $B_s(t)\to X_1\,X_2$ can be written as \cite{dighethank}
\beq\lab{e10}
f(\theta,\varphi,\psi;t)=\sum_kb^{(k)}(t)g^{(k)}(\theta,\varphi,\psi),
\eeq
where we have denoted the angles describing the kinematics of the decay
products of $X_1$ and $X_2$ generically by $\theta$, $\varphi$ and
$\psi$. Note that we have to deal in general with an arbitrary number of 
such angles. For quasi two body modes $B_s(t)\to X_1\,X_2$ into final states
that are admixtures of CP-even and CP-odd configurations, 
the observables $b^{(k)}(t)$ describing the
time-evolution of the angular distribution (\ref{e10}) can be expressed 
in terms of real or imaginary parts of bilinear combinations of decay 
amplitudes having the same structure as (\ref{e5}). The angular
distribution for the {\it tagged} CP-conjugate decay $\bsbar(t)\to X_1\,X_2$ 
on the other hand is given by
\beq\lab{e11}
\bar f(\theta,\varphi,\psi;t)=\sum_k\bar b^{(k)}(t)g^{(k)}(\theta,
\varphi,\psi),
\eeq
where the observables $\bar b^{(k)}(t)$ are related correspondingly to
real or imaginary parts of bilinear combinations like (\ref{e8}).
Since the states $X_1\,X_2$ resulting from the $B_s$ and $\bsbar$ decays 
are equal, we use the same generic angles $\theta$, $\varphi$ and $\psi$ 
to describe the angular distributions of their decay products. Within
our formalism the effects of CP transformations relating $B_s(t)\to
\left(X_1\,X_2\right)_{f,\tilde f}$ and $\overline{B_s}(t)\to\left(X_1\,
X_2\right)_{f,\tilde f}$ are taken into account already by the
CP eigenvalues $\etacptilde$ and $\etacp$ appearing in (\ref{e5})
and (\ref{e8}) and do not affect $g^{(k)}(\theta,\varphi,\psi)$. 
Therefore the same functions $g^{(k)}(\theta,\varphi,\psi)$ are
present in (\ref{e10}) and (\ref{e11}). 

\vspace{0.3cm}

The main focus of this paper are {\it untagged} rates, where one does not
distinguish between initially present $B_s$ and $\bsbar$ mesons. Such 
studies are obviously much more efficient from an experimental point 
of view than tagged analyses. In the distant future it will
become feasible to collect also {\it tagged} $B_s$ data samples and to 
resolve the rapid oscillatory $\Delta mt$--terms. Then Eqs.~(\ref{e5})
and (\ref{e8}) describing the corresponding observables should turn out
to be very useful. 

Combining (\ref{e10}) and (\ref{e11}) we find that the {\it untagged} 
angular distribution takes the form 
\beq\lab{e12}
[f(\theta,\varphi,\psi;t)]\equiv\bar f(\theta,\varphi,\psi;t)+ 
f(\theta,\varphi,\psi;t)=\sum_k\left[\bar b^{(k)}(t)+b^{(k)}(t)\right]
g^{(k)}(\theta,\varphi,\psi).
\eeq
As we will see in a moment, interesting CP-violating effects show up
in this untagged rate, if the width difference $\Delta
\Gamma$ is sizable. The time-evolution of the relevant observables $\left[\bar 
b^{(k)}(t)+b^{(k)}(t)\right]$ behaves as the real or imaginary 
parts of 
\bea
\lefteqn{\left[\aftilde(t)\,\af(t)\right]\equiv\aftildebar(t)\,\afbar(t)+
\aftilde(t)\,\af(t)=\frac{1}{2}\langle\xxtilde|
\heff|B_s\rangle^\ast\langle\xx|
\heff|B_s\rangle}\nonumber\\
&&\times\left[\left(1+\etacptilde\,\etacp\xiftilde^\ast\,\xif
\right)\left(e^{-\Gamma_Lt}+e^{-\Gamma_Ht}\right)+\left(\etacptilde\,
\xiftilde^\ast+\etacp\,\xif\right)\left(e^{-\Gamma_Lt}-
e^{-\Gamma_Ht}\right)\right].\lab{e13}
\eea
In order to calculate this equation, we have combined (\ref{e5}) 
with (\ref{e8}) and have moreover taken into account explicitly 
the time-dependences of (\ref{e6a}) and (\ref{e6b}). We can
distinguish between the following special cases:
\begin{itemize}
\item $\tilde f=f$:
\bea
\lefteqn{\left[|\af(t)|^2\right]=\frac{1}{2}|\langle\xx|\heff|B_s\rangle|^2}
\lab{es1}\\
&&\times\left[\left(1+|\xif|^2\right)\left(e^{-\Gamma_Lt}+
e^{-\Gamma_Ht}\right)+2\,\etacp\mbox{\,Re}(\xif)\left(e^{-\Gamma_Lt}-
e^{-\Gamma_Ht}\right)\right]\nonumber
\eea
\item $\tilde f\not=f$ and $\etacptilde=\etacp$:
\bea
\lefteqn{\left[\aftilde(t)\,\af(t)\right]=\frac{1}{2}\langle\xxtilde|
\heff|B_s\rangle^\ast\langle\xx|
\heff|B_s\rangle}\lab{es2}\\
&&\times\left[\left(1+\xiftilde^\ast\,\xif\right)
\left(e^{-\Gamma_Lt}+e^{-\Gamma_Ht}\right)+\etacp\,\left(\xiftilde^\ast
+\xif\right)\left(e^{-\Gamma_Lt}-e^{-\Gamma_Ht}\right)\right]\nonumber
\eea
\item $\tilde f\not=f$ and $\etacptilde=-\etacp$:
\bea
\lefteqn{\left[\aftilde(t)\,\af(t)\right]=\frac{1}{2}\langle\xxtilde|
\heff|B_s\rangle^\ast\langle\xx|
\heff|B_s\rangle}\lab{es3}\\
&&\times\left[\left(1-\xiftilde^\ast\,\xif\right)
\left(e^{-\Gamma_Lt}+e^{-\Gamma_Ht}\right)-\etacp\,\left(\xiftilde^\ast
-\xif\right)\left(e^{-\Gamma_Lt}-e^{-\Gamma_Ht}\right)\right].\nonumber
\eea
\end{itemize}

As advertised, the rapidly oscillating $\Delta mt$--terms 
cancel in the untagged combinations described by (\ref{e13}). 
While the time-dependence of (\ref{es1}) was given in \cite{dunietz},
the explicit time-dependences of (\ref{es2}) and (\ref{es3}) have
not been given previously. They play an important role for the untagged
angular distribution (\ref{e12}).

\section{Dominance of a single CKM amplitude}\lab{dom}

If we look at expression (\ref{e7}), we observe that $\xif$ and
$\xiftilde$ suffer in general from large hadronic uncertainties.
However, if the unmixed $B_s\to X_1\,X_2$ amplitude is dominated 
by a single CKM amplitude proportional to a CKM factor $v_j^{(r)}$, 
the unknown hadronic matrix elements cancel in (\ref{e7}) and both 
$\xiftilde$ and $\xif$ take the simple form
\beq\lab{e15}
\xiftilde=\xif=e^{2i\phi_j^{(r)}},
\eeq
where $\phi_j^{(r)}\equiv\left(\mbox{arg}(V_{jr}^\ast V_{jb})-
\mbox{arg}(V_{ts}^\ast V_{tb})\right)$
is a CP-violating weak phase consisting of the corresponding decay and
$B_s-\overline{B_s}$ mixing phase. Consequently, in that very
important special case, (\ref{e13}) simplifies to
\bea
\lefteqn{\left[\aftilde(t)\,\af(t)\right]=\frac{1}{2}
|\langle\xxtilde|\heff|B_s\rangle\langle\xx|\heff|B_s\rangle|
e^{i(\delta_f-\delta_{\tilde f})}}\lab{e16}\\
&&\times\left[\left(1+\etacptilde\,\etacp\right)
\left(e^{-\Gamma_Lt}+e^{-\Gamma_Ht}\right)
+\left(\etacptilde\,e^{-2i\phi^{(r)}_j}+\etacp\,e^{2i\phi^{(r)}_j}\right)
\left(e^{-\Gamma_Lt}-e^{-\Gamma_Ht}\right)\right],\nonumber
\eea
where $\delta_f$ and $\delta_{\tilde f}$ are {\it CP-conserving}
strong phases. They are induced through strong final state interaction
processes and are defined by
\bea
\langle\xx|\heff|B_s\rangle&=&e^{+i\delta_f}e^{-i\phi_j^{(r)}}\lab{e18}\\
\langle\xxtilde|\heff|B_s\rangle^\ast
&=&e^{-i\delta_{\tilde f}}e^{+i\phi_j^{(r)}}.\lab{e19}
\eea
Note that the structure of (\ref{e18}) and (\ref{e19}), which is 
essentially due to the fact that the unmixed 
$B_s\to X_1\,X_2$ amplitude is dominated by a single weak 
amplitude, implies that the weak phase factors $e^{-i\phi_j^{(r)}}$ and 
$e^{+i\phi_j^{(r)}}$ cancelled each other in (\ref{e16}) and 
that only the strong phases play a role as an overall phase in this equation. 
We would like to emphasize that such a simple behavior is not present 
in the general case where more than one weak amplitude is present.  

The time-evolution of (\ref{e16}) depends only on $\cos2\phi^{(r)}_j$
and $\sin2\phi^{(r)}_j$, since we have only to deal with the following
two cases:
\begin{itemize}
\item $\etacptilde=\etacp$:
\bea
\lefteqn{\left[\aftilde(t)\,\af(t)\right]=|\langle\xxtilde|
\heff|B_s\rangle\langle\xx|
\heff|B_s\rangle|e^{i(\delta_f-\delta_{\tilde f})}}\lab{es4}\\
&&\times\left[\left(e^{-\Gamma_Lt}+e^{-\Gamma_Ht}\right)+
\etacp\,\left(e^{-\Gamma_Lt}-e^{-\Gamma_Ht}\right)\cos2\phi^{(r)}_j
\right]\nonumber
\eea
\item $\etacptilde=-\etacp$:
\bea
\lefteqn{\left[\aftilde(t)\,\af(t)\right]=}\lab{es5}\\
&&|\langle\xxtilde|
\heff|B_s\rangle\langle\xx|\heff|B_s\rangle|
e^{i(\delta_f-\delta_{\tilde f})}
\,i\,\etacp\,\left(e^{-\Gamma_Lt}-e^{-\Gamma_Ht}\right)
\sin2\phi^{(r)}_j.\nonumber
\eea
\end{itemize}

Whereas the structure of (\ref{es4}), in particular the $\cos2\phi_j^{(r)}$
term, has already been discussed for $\tilde f=f$ in \cite{dunietz}, to the
best of our knowledge it has not been pointed out so far that untagged
data samples of angular distributions for certain non-leptonic $B_s$ 
decays allow also a determination of $\sin2\phi_j^{(r)}$ with the help of
(\ref{es5}). These $\sin2\phi_j^{(r)}$ terms play an important role if the
weak phase $\phi_j^{(r)}$ is small. The point is that $\sin2\phi_j^{(r)}$
is proportional to $\phi_j^{(r)}$ in that case, while $\cos2\phi_j^{(r)}
=1+\order\left(\phi_j^{(r)2}\right)$. Consequently we obtain up to 
terms of $\order\left(\phi_j^{(r)2}\right)$:
\begin{itemize}
\item $\etacptilde=\etacp=+1$:
\beq\lab{e20}
\left[\aftilde(t)\af(t)\right]=2|\langle\xxtilde|\heff|B_s\rangle
\langle\xx|\heff|B_s\rangle|e^{i(\delta_f-\delta_{\tilde f})}e^{-\Gamma_Lt}
\eeq
\item $\etacptilde=\etacp=-1$:
\beq\lab{e21}
\left[\aftilde(t)\af(t)\right]=2|\langle\xxtilde|\heff|B_s\rangle
\langle\xx|\heff|B_s\rangle|e^{i(\delta_f-\delta_{\tilde f})}e^{-\Gamma_Ht}
\eeq
\item $\etacptilde=-\etacp$:
\bea
\lefteqn{\left[\aftilde(t)\af(t)\right]=}\lab{e22}\\
&&2\,i\,\etacp\,
|\langle\xxtilde|\heff|B_s\rangle\langle\xx|
\heff|B_s\rangle|e^{i(\delta_f-\delta_{\tilde f})}
\left(e^{-\Gamma_Lt}-e^{-\Gamma_Ht}\right)\phi_j^{(r)}.\nonumber
\eea
\end{itemize}
We observe that only the mixed combination (\ref{e22}) is sensitive, i.e.\
proportional, to the small phase $\phi_j^{(r)}$ and allows an extraction 
of this quantity. These considerations have an interesting phenomenological
application as we will see in the following section.

\section{The ``gold-plated'' transitions $B_s\to D_s^{\ast+}\,D_s^{\ast-}$
and $B_s\to J/\psi\,\phi$ to extract the Wolfenstein parameter 
$\eta$}\lab{gold}

Concerning the dominance of a single CKM amplitude, in analogy to 
$B_d\to J/\psi\, K_S$ measuring $\sin2\beta$ to excellent accuracy
\cite{bs} ($\beta$ is another angle of the unitarity triangle \cite{ut}),
the ``gold-plated'' modes are $B_s$ decays caused by $\bar b\to\bar cc\bar s$
quark-level transitions. The corresponding exclusive modes relevant for
our discussion are $B_s\to D_s^{\ast +}(\to D_s^+\gamma)\,D_s^{\ast-}
(\to D_s^-\gamma)$ and $B_s\to J/\psi(\to l^+l^-)\,\phi(\to K^+K^-)$. They
are dominated to an excellent accuracy by the CKM amplitudes proportional
to $v_c^{(s)}=V_{cs}^\ast V_{cb}$. Therefore the corresponding weak phase
$\phi_j^{(r)}$ defined after (\ref{e15}) is related to elements of the 
CKM matrix
\cite{km} through
\beq\lab{e23}
\phi_c^{(s)}=[\mbox{arg}(V_{cs}^\ast V_{cb})-
\mbox{arg}(V_{ts}^\ast V_{tb})].
\eeq
At leading order in the Wolfenstein expansion \cite{wolf} this phase 
vanishes. In order to obtain a non-vanishing result, we have to take into 
account higher order terms in the Wolfenstein parameter $\lambda=
\sin\theta_{\mbox{{\scriptsize C}}}=0.22$ (for a treatment of such terms 
see e.g.\ \cite{wolf,blo}) yielding \cite{thesisd,buras1}
\beq\lab{e24}
\phi_c^{(s)}=\lambda^2\eta=\order(0.015).
\eeq
Consequently the small weak phase $\phi_c^{(s)}$ measures
simply the CKM parameter $\eta$ \mbox{\cite{wolf,thesisd,buras1}}.
 
Another interesting interpretation of (\ref{e23}) is the fact that it is 
related to one angle in a rather squashed (and therefore ``unpopular'') 
unitarity triangle \cite{akl}. Other useful expressions for (\ref{e23}) 
can be found in~\cite{snowmass93}. If we use the CKM factor $R_b$ defined 
by (\ref{Rb}) as an additional input, $\eta$ fixes the notoriously 
difficult to measure angle $\gamma$ of the unitarity triangle 
\cite{snowmass93}. That input allows, however, also a determination 
of $\gamma$ (or of the Wolfenstein parameter $\eta$) from the
mixing-induced CP-violating asymmetry arising in $B_d\to J/\psi\,K_S$
measuring $\sin2\beta$. Comparing these two results for $\gamma$
(or $\eta$), an interesting test whether the phases in 
$B_d-\overline{B_d}$ and $B_s-\overline{B_s}$ mixing are
indeed described by the Standard Model can be performed. 

The extraction of the weak phase Eq.~(32) from $B_s \rightarrow J/\psi\,
\phi$, $D^{*+}_s\,D^{*-}_s$, etc.\ is not as clean as that of $\beta$ from 
$B_d\to J/\psi\,K_S$. The reason is that although the contributions 
to the unmixed amplitudes proportional to $V^*_{ub} V_{us}$ are similarly 
suppressed in both cases, their importance is enhanced by the smallness 
of $\phi^{(s)}_c$ versus $\beta$~\cite{kayserlprivate}.

Given that $\phi_c^{(s)}$ is small, we see that 
(\ref{e20})-(\ref{e22}) apply to an excellent approximation to the 
exclusive channels $B_s\to D_s^{\ast +}(\to D_s^+\gamma)\,D_s^{\ast-}
(\to D_s^-\gamma)$ and $B_s\to J/\psi(\to l^+l^-)\,\phi(\to K^+K^-)$,
i.e.\ to $X_1\,X_2\in\{D_s^{\ast+}D_s^{\ast-},J/\psi\,\phi\}$.
Whereas the angular distribution of the latter process has been  
derived in \cite{ddlr}, a follow-up note \cite{ddf1} not only  
examines the angular distributions for both processes but also  
discusses an efficient method for determining the relevant  
observables -- the {\it moment analysis} \cite{dqstl} -- and 
predicts these observables, thereby allowing comparisons with future 
experimental data.

The combination (\ref{e22}) enters the {\it untagged} angular
distribution in the form
\bea
\lefteqn{\mbox{Im}\left\{\left[\aftilde(t)\af(t)\right]\right\}=}\lab{e25}\\
&&-2|\langle\xxtilde|\heff|B_s\rangle\langle\xx|
\heff|B_s\rangle|\cos(\delta_f-\delta_{\tilde f})
\left(e^{-\Gamma_Lt}-e^{-\Gamma_Ht}\right)\phi_c^{(s)},\nonumber
\eea
where $\tilde f\in\{\,\parallel\,,\,0\,\}$ and $f=\,\perp$ denote linear
polarization states \cite{rosner, ddlr}. In order to determine the
weak phase $\phi_c^{(s)}$ from (\ref{e25}), we have to know both
$|\langle\xxtilde|\heff|B_s\rangle|$, $|\langle\xx|\heff|B_s\rangle|$
and the strong phase differences $\delta_f-\delta_{\tilde f}$. 
Whereas the former quantities can be determined straightforwardly 
from
\bea
\left[|\af(t)|^2\right]&=&2|\langle\xx|\heff|B_s\rangle|^2e^{-\Gamma_Lt}
\qquad(f\in\{\parallel,0\})\\
\left[|A_\perp(t)|^2\right]&=&2|\langle(X_1\,X_2)_\perp|\heff|B_s\rangle|^2
e^{-\Gamma_Ht},
\eea
the latter ones can be obtained by combining the ratio of (\ref{e25}) 
for $\tilde f=\,\parallel$ and $\tilde f=0$ given by
\beq
\frac{\mbox{Im}\{[A_\parallel^\ast(t)A_\perp(t)]\}}
{\mbox{Im}\{[A_0^\ast(t)A_\perp(t)]\}}=
\frac{|\langle(X_1\,X_2)_\parallel|\heff|B_s\rangle|}
{|\langle(X_1\,X_2)_0|\heff|B_s\rangle|}
\frac{\cos(\delta_\perp-\delta_\parallel)}{\cos(\delta_\perp-\delta_0)}
\eeq
with the term of the untagged angular distribution corresponding 
to \cite{ddlr,ddf1}
\beq
\mbox{\,Re}\left\{\left[A_0^\ast(t)A_\parallel(t)\right]\right\}=
2|\langle(X_1\,X_2)_0|\heff|B_s\rangle
\langle(X_1\,X_2)_\parallel|\heff|B_s\rangle|
\cos(\delta_\parallel-\delta_0)e^{-\Gamma_Lt}.
\eeq
Consequently the angular distributions for the {\it untagged} 
$B_s\to D_s^{\ast +}(\to D_s^+\gamma)\,D_s^{\ast-}
(\to D_s^-\gamma)$ and $B_s\to J/\psi(\to l^+l^-)\,\phi(\to K^+K^-)$ modes
allow a determination of the weak phase $\phi_c^{(s)}$.

The rather complicated extraction of the strong phase differences 
$\delta_f-\delta_{\tilde f}$ outlined above, which is needed to 
accomplish this task, can, however, be simplified
considerably by making an additional assumption. In the
case of the color-allowed channel $B_s\to D_s^{\ast +}\,D_s^{\ast-}$
the {\it factorization hypothesis} \cite{fact1,fact2}, which can be 
justified to some extent within the 
$1/N_{\mbox{{\scriptsize C}}}$--expansion \cite{bgr}, predicts
rather reliably that the strong phase shifts are $0~\mbox{mod}~\pi$.
This prediction for the strong phases can be tested experimentally by
investigating the angular correlations for the $SU(3)$-related modes
$B_{u,d}\to D_s^{\ast+}\overline{D^\ast}_{u,d}$.
Since $B_s\to J/\psi\,\phi$ is on the other hand a color-suppressed
transition, the validity of the factorization approach is very
doubtful in this case \cite{bjorken}. However, flavor $SU(3)$ 
symmetry of strong interactions is probably a good working assumption 
and can be used to determine the hadronization dynamics of 
$B_s\to J/\psi\,\phi$, in particular the strong phase differences
$\delta_f-\delta_{\tilde f}$, from an analysis of the
$SU(3)$-related $B\to J/\psi\, K^\ast$ modes \cite{ddf1,dqstl}.
These strategies should be very helpful to constrain $\phi_c^{(s)}$
with more limited statistics. 

Whereas one expects $\Gamma_H<\Gamma_L$ and a small value of $\phi_c^{(s)}$
within the Standard Model, that need not to be the case in many scenarios
for ``New Physics'' beyond the Standard Model (see e.g.\ 
\cite{grossman}). The untagged data samples described by (\ref{es4})
and (\ref{es5}) allow then only the extraction of $\cos 2\phi_c^{(s)}$
and $\sin 2\phi_c^{(s)}$ up to some discrete ambiguities. In particular
they do not allow the determination of the sign of $\Delta\Gamma$ 
which could give us hints to physics beyond the Standard Model. This
feature is simply due to the fact that we cannot decide which decay 
width is $\Gamma_L$ and $\Gamma_H$, respectively, since we do not know the
sign of $\Delta\Gamma$. Using, however, in addition the time-dependences
of {\it tagged} data samples, $\sin 2\phi_c^{(s)}$ can be extracted and
the discrete ambiguities are resolved. With the help of the observables 
corresponding to (\ref{es5}) even the sign of $\Delta\Gamma$ can then be 
extracted, which was missed in a recent note \cite{grossman}.  In general,
the ambiguities encountered in studies of untagged data samples are 
resolved by incorporating the additional information available from 
$\Delta m t-$oscillations.
 
\section{A determination of $\gamma$ using untagged data samples
of $B_s\to K^{\ast+}\,K^{\ast-}$ and $B_s\to K^{\ast 0}\,
\overline{K^{\ast0}}$}\lab{gam}

After our discussion of some exclusive $\bar b\to\bar cc\bar s$ transitions
and a brief excursion to ``New Physics'' in the previous section let 
us now consider the $\bar b\to\bar uu\bar s$ 
decay $B_s\to K^{\ast +}(\to\pi K)\,K^{\ast -}(\to\pi 
\overline{K})$ and investigate what can be learned  
from {\it untagged} measurements of its angular distribution. Because of 
the special CKM-structure of the $\bar b\to\bar s$ 
penguins~\cite{bf}, their contributions to $B_s\to K^{\ast+}\,K^{\ast-}$
can be written in the form 
\beq\lab{pens}
P_f'=-|P_f'|e^{i\deltap}e^{i\pi},
\eeq
where $f$ denotes final state configurations of $K^{\ast+}\,K^{\ast-}$
with CP eigenvalue $\etacp$ (see (\ref{e3})), $\deltap$ are CP-conserving
strong phases, the CP-violating weak phase has the numerical value
of $\pi$ and the minus sign is due to our definition of meson states 
which is similar to the conventions applied in \cite{ghlrSU3}. 

The penguin contributions include not only penguins with internal top-quark 
exchanges, but also those with internal up- and charm-quarks \cite{bf}.
Rescattering processes are included by definition in the penguin 
amplitude $P_f'$. For example, the process $B_s\to\{D_s^{\ast+}\,
D_s^{\ast-}\}\to K^{\ast+}K^{\ast-}$ (see e.g.\ \cite{kamal}) is related
to penguin topologies with charm-quarks running in the loops as can be seen 
easily by drawing the corresponding Feynman diagrams. Although such
rescattering processes may affect $|P_f'|$ and $\deltap$, they do not
modify the weak phase in (\ref{pens}).

On the other hand the contributions of the current-current operators 
appearing in the low energy effective 
Hamiltonian (\ref{ham}), which are color-allowed in the case of 
$B_s\to K^{\ast+}\,K^{\ast-}$, have the structure 
\beq
T_f'=-|T_f'|e^{i\delta_{T'}^f}e^{i\gamma},
\eeq
where $\delta_{T'}^f$ is again a CP-conserving strong phase. 
Consequently, combining these considerations, we obtain the following
transition matrix element for $B_s\to\xx$ with 
$X_1\,X_2=K^{\ast+}\,K^{\ast-}$:
\beq\lab{hama}
\langle\xx|\heff|B_s\rangle=|\pf|e^{i\deltap}\left[1-\rf e^{i\gamma}\right],
\eeq
where
\beq
\rf\equiv\frac{|T_f'|}{|\pf|}e^{i(\delta_{T'}^f-\deltap)}.
\eeq
Hence the quantitiy $\xif$ defined through (\ref{e7}) is given by
\beq
\xif=\frac{1-\rf e^{-i\gamma}}{1-\rf e^{+i\gamma}}.
\eeq
Following the plausible hierarchy of decay amplitudes introduced in
\cite{ghlrSU3}, we expect that penguins play -- in analogy to
$B_s\to K^+\,K^-$ \cite{PAPI,PAPIII} -- the dominant role in
$B_s\to K^{\ast+}\,K^{\ast-}$. 

To evaluate the time-evolution of the observables of the untagged
angular distribution corresponding to real or imaginary parts of
(\ref{e13}), we need $1\pm\xiftilde^\ast\,\xif$ and $\xiftilde^\ast
\pm\xif$ which are given by 
\bea
1+\xiftilde^\ast\,\xif&=&\frac{2}{\nff}\left[1-\left(\rftilde+\rf\right)
\cos\gamma+\rftilde\,\rf\right]\\
1-\xiftilde^\ast\,\xif&=&i\,\frac{2}{\nff}\left(\rftilde-\rf\right)
\sin\gamma
\eea
and
\bea
\xiftilde^\ast+\xif&=&\frac{2}{\nff}\left[1-\left(\rftilde+\rf\right)
\cos\gamma+\rftilde\,\rf\,\cos2\gamma\right]\\
\xiftilde^\ast-\xif&=&-\,i\,\frac{2}{\nff}\left[\rftilde+\rf
-2\rftilde\,\rf\,\cos\gamma\right]\sin\gamma,
\eea
respectively, where
\beq
\nff\equiv1-\rftilde\,e^{-i\gamma}-\rf\,e^{i\gamma}+\rftilde\,\rf.
\eeq
These combinations of $\xiftilde^\ast$ and $\xif$ are multiplied 
in (\ref{e13}) by 
\beq
\langle\xxtilde|\heff|B_s\rangle^\ast\langle\xx|\heff|B_s\rangle=|\pftilde|\,
|\pf|e^{i(\deltap-\deltaptilde)}\nff.
\eeq
Here we have used the expression (\ref{hama}) to calculate this 
product of hadronic 
matrix elements, which -- in contrast to the case where a single
CKM amplitude dominates (see the cautious remark after (\ref{e19})) --
depends also on the weak phase $\gamma$ through $\nff$. However, these
factors cancel in (\ref{e13}) so that we finally arrive at the
following set of equations describing $B_s\to (K^{\ast+}\,K^{\ast-})_f$:
\begin{itemize}
\item $\etacptilde=\etacp=+1$:
\bea
\lefteqn{\left[\aftilde(t)\af(t)\right]=2\,|\pftilde|\,|\pf|e^{i(\deltap-
\deltaptilde)}}\lab{ekk1}\\
&&\times\left[\left\{1-\left(\rftilde+\rf\right)\cos\gamma+\rftilde\,
\rf\,\cos^2\gamma\right\}e^{-\Gamma_Lt}+\rftilde\,\rf\,\sin^2\gamma\,
e^{-\Gamma_Ht}\right]\nonumber
\eea
\item $\etacptilde=\etacp=-1$:
\bea
\lefteqn{\left[\aftilde(t)\af(t)\right]=2\,|\pftilde|\,|\pf|e^{i(\deltap-
\deltaptilde)}}\lab{ekk2}\\
&&\times\left[\left\{1-\left(\rftilde+\rf\right)\cos\gamma+\rftilde\,
\rf\,\cos^2\gamma\right\}e^{-\Gamma_Ht}+\rftilde\,\rf\,\sin^2\gamma\,
e^{-\Gamma_Lt}\right]\nonumber
\eea
\item $\etacptilde=-\etacp=+1$:
\bea
\lefteqn{\left[\aftilde(t)\af(t)\right]=i\,2\,|\pftilde|\,|\pf|
e^{i(\deltap-\deltaptilde)}}\lab{ekk3}\\
&&\times\left[\rftilde\,e^{-\Gamma_Ht}-\rf\,e^{-\Gamma_Lt}+
\rftilde\,\rf\,\left(e^{-\Gamma_Lt}-e^{-\Gamma_Ht}\right)\cos\gamma\right]
\sin\gamma.\nonumber
\eea
\end{itemize}
The structure of these equations, which are valid exactly, is much more
complicated than that of (\ref{e20})-(\ref{e22}) where a single CKM
amplitude dominates to an excellent accuracy. Note that a measurement
of either the $e^{-\Gamma_Ht}$ or $e^{-\Gamma_Lt}$ terms in (\ref{ekk1})
and (\ref{ekk2}), respectively, or of non-vanishing observables 
corresponding to (\ref{ekk3}) would give unambiguous evidence for a 
non-vanishing value of $\sin\gamma$. 

A determination of $\gamma$ is possible if one measures in addition the
time-dependent {\it untagged} angular distribution for 
$B_s\to K^{\ast0}\,\overline{K^{\ast0}}$ which is a pure penguin-induced
$\bar b\to\bar sd\bar d$ transition. Its time-evolution can be obtained
from (\ref{ekk1})-(\ref{ekk3}) by setting $r_{\tilde f}=\rf=0$ and 
depends only on the hadronization dynamics of the penguin operators. 

There are two classes of penguin topologies as we have already noted
briefly after~(\ref{e2c}): QCD and electroweak penguins
originating from strong and electroweak interactions, respectively. In
contrast to na\"\i ve expectations, the contributions of electroweak 
penguin operators may play an important role in certain non-leptonic 
$B$-meson decays because of the presence of the {\it heavy} top-quark
\cite{rfewp1,rfewp} (see also \cite{dh}-\cite{ghlrEWP}). However, 
in the case of the $B_s\to K^\ast \overline{K^\ast}$ transitions considered 
in this section, these contributions are color-suppressed and play only
a minor role compared to those of the dominant QCD penguin operators.  

If we neglect these electroweak penguin contributions, which has not been 
done in the formulae given above and should be a good approximation in 
our case, and use furthermore the $SU(2)$ isospin symmetry of strong 
interactions, the $B_s\to K^{\ast0}\,\overline{K^{\ast0}}$ observables 
can be related to the $B_s\to K^{\ast+}\,K^{\ast-}$ case. 
In terms of linear polarization states \cite{rosner}, these 
observables fix $|P_0'|$, $|P_\parallel'|$, 
$|P_\perp'|$ and $\cos(\delta_{P'}^0-\delta_{P'}^\parallel)$. Since the
overall normalizations of the untagged $B_s\to K^{\ast+}\,K^{\ast-}$
observables can be determined this way, the $e^{-\Gamma_Lt}$ and
$e^{-\Gamma_Ht}$ pieces of the observables $[|A_0(t)|^2]$, $[|A_\parallel
(t)|^2]$ and Re$\{[A_0^\ast(t)\,A_\parallel(t)]\}$ (see (\ref{ekk1}))  
allow another extraction of the CKM angle $\gamma$. The remaining
observables can be used to resolve possible discrete
ambiguities. Needless to say, also the quantities $r_f$ and the QCD
penguin amplitudes $P_f$ are of particular interest since they provide
insights into the hadronization dynamics of the QCD penguins. 
A detailed analysis of the decays $B_s\to K^{\ast+}\,K^{\ast-}$ and 
$B_s\to K^{\ast0}\,\overline{K^{\ast0}}$ is presented in \cite{ddf2}, 
where also the angular distributions are given explicitly.

Another interesting application of (\ref{ekk1}) is associated with the
decays $B_s\to K^+ K^-$ and $B_s\to K^0\,\overline{K^0}$. Using again
the $SU(2)$ isospin symmetry of strong interactions to relate their
QCD penguin contributions (electroweak penguin contributions are once 
more color-suppressed and are hence very small), the time-dependent 
{\it untagged} rates for these modes evolve as
\beq\lab{eA}
\left[|A(t)|^2\right]=2\,|P'|^2\left[(1-2\,|r|\cos\rho\,\cos\gamma
+|r|^2\cos^2\gamma)e^{-\Gamma_Lt}+|r|^2\sin^2\gamma\, e^{-\Gamma_Ht}
\right]
\eeq
and
\beq
\left[|A(t)|^2\right]=2\,|P'|^2\,e^{-\Gamma_Lt},
\eeq
respectively, where we have used
\beq
r\equiv|r|e^{i\rho}.
\eeq
Here $\rho$ is a CP-conserving strong phase and $|r|=|T'|/|P'|$. 
In general, there are a lot fewer observables in  
``pseudoscalar-pseudoscalar'' cases than in ``vector-vector'' cases. In 
particular there is no observable corresponding to Re$\{[A^\ast_0(t)\,
A_\parallel(t)]\}$. We therefore need some additional input in order to 
extract $\gamma$ from (\ref{eA}).
That is provided by the $SU(3)$ flavor symmetry of strong
interactions. If we neglect the color-suppressed current-current
contributions to $B^+\to\pi^+\pi^0$, which are expected to be
suppressed relative to the color-allowed contributions by a factor
of $\order(0.2)$, this symmetry yields \cite{ghlrSU3}
\beq\lab{eT}
|T'|\approx\lambda\,\frac{f_K}{f_\pi}\,\sqrt{2}\,|A(B^+\to\pi^+\pi^0)|,
\eeq
where $\lambda$ is the Wolfenstein parameter \cite{wolf}, $f_K$ and
$f_\pi$ are the $K$- and $\pi$-meson decay constants, respectively,
and $A(B^+\to\pi^+\pi^0)$ denotes the appropriately normalized 
$B^+\to\pi^+\pi^0$ decay amplitude. Since $|P'|$ is known from
$B_s\to K^0\,\overline{K^0}$, the quantity $|r|$ can be estimated 
with the help of (\ref{eT}) and allows the extraction of $\gamma$
from the part of (\ref{eA}) evolving with the exponent $e^{-\Gamma_Ht}$.
Using in addition the piece evolving with $e^{-\Gamma_Lt}$ the strong
phase $\rho$ can also be determined up to certain discrete ambiguities.
Since one expects $|r|=\order(0.2)$ \cite{ghlrSU3,PAPI,PAPIII}, it may be
difficult to measure the $e^{-\Gamma_Ht}$ contribution to (\ref{eA})
which is proportional to $|r|^2$. The value of $\gamma$ and the observable
$r$ estimated that way could be used as an input to determine
electroweak penguin amplitudes by measuring in addition the branching
ratios BR$(B^+\to\pi^0K^+)$, BR$(B^-\to\pi^0K^-)$ and 
BR$(B^+\to\pi^+K^0)=\mbox{BR}(B^-\to\pi^-\overline{K^0})$ as has been 
proposed in \cite{PAPI}.

Let us finally note that (\ref{ekk1})-(\ref{ekk3}) apply also to the
mode $B_s\to\rho^0\phi$, if we perform the replacements
\bea
|P_f'|&\to&|P_f^{'EW}|\nonumber\\
\delta_{P'}^f&\to&\delta_{EWP'}^f\\
r_f&\to&\frac{|C_f'|}{|P_f^{'EW}|}\exp\left[i\left(\delta_{C'}^f-
\delta_{EWP'}^f\right)\right]\nonumber,
\eea
where $C_f'$ denotes color-suppressed contributions of the 
current-current operators and $|P_f^{'EW}|$, $\delta_{EWP'}^f$ are
related to color-allowed contributions of electroweak
penguin operators. Similar to the situation arising in $B_s\to\pi^0\phi$,
which has been discussed in \cite{rfewp} 
(see also \mbox{\cite{dht,dy,ghlrEWP}}),
we expect that this decay is dominated by electroweak penguins. 
Consequently its {\it untagged} angular distribution may inform us 
about the physics of the corresponding operators. In respect of 
controlling electroweak penguins in a quantitative way by using 
$SU(3)$ relations among $B\to\pi K$ decay amplitudes \cite{PAPI},
the CKM angle $\gamma$ is a central input. Therefore the new strategies
to extract this angle in a rather clean way from {\it untagged} $B_s$ data
samples presented in Sections~\ref{gold} and \ref{gam} are also very
helpful to accomplish this ambitious task.

\section{Summary}\lab{sum}

We have calculated the time-evolutions of angular distributions for
$B_s$ decays into final states that are admixtures of different CP 
eigenstates. Interestingly, due to the expected perceptible 
$B_s-\overline{B_s}$ lifetime difference, the corresponding 
observables may allow the extraction
of CKM phases even in the {\it untagged} case where one does not
distinguish between initially present $B_s$ and $\overline{B_s}$
mesons. As we have demonstrated in this paper, such studies of the
exclusive $\bar b\to\bar cc\bar s$ modes $B_s\to D_s^{\ast +}\,
D_s^{\ast -}$ and $B_s\to J/\psi\,\phi$, which are dominated to an
excellent approximation by a single CKM amplitude, allow a  
determination of the Wolfenstein parameter $\eta$ thereby fixing the
height of the usual unitarity triangle. Using the CKM factor $R_b
\propto |V_{ub}|/|V_{cb}|$ as an additional input, $\gamma$ can be 
determined both from $\eta$ and from mixing-induced CP-violation
in $B_d\to J/\psi\, K_S$ measuring $\sin 2\beta$. A comparison
of these two results for $\gamma$ determined from $B_s$ and $B_d$
decays, respectively, would allow an interesting test whether the
corresponding mixing phases are described by the Standard Model.

If we apply the $SU(2)$ isospin symmetry of strong interactions to
relate the QCD penguin contributions to the $\bar b\to\bar uu\bar s$
mode $B_s\to K^{\ast+}\,K^{\ast-}$ and to the $\bar b\to\bar sd\bar d$
transition $B_s\to K^{\ast0}\,
\overline{K^{\ast0}}$, which should play the dominant role there,
another extraction of $\gamma$ is possible from {\it untagged} 
measurements of their angular distributions. Substituting the
relevant variables appropriately, the results derived for $B_s\to
K^{\ast+}K^{\ast-}$ apply also to $B_s\to\rho^0\phi$ which is expected
to be dominated by electroweak penguin operators. 

We will come back to these decays in separate forthcoming publications 
\cite{ddf1,ddf2}. The case of $B_s$ decays into final states that 
are not admixtures of different CP eigenstates but only of different 
parity eigenstates is outlined in \cite{df2}. There we discuss how
angular correlations for untagged $B_s$ decays governed by 
$\bar b\to\bar cu\bar s$ quark-level transitions allow also a  
determination of the CKM angle $\gamma$.

\section*{Acknowledgments}
We are very grateful to Helen Quinn for a critical reading of the
manuscript. R.F.\ would like to thank Helen Quinn for interesting
discussions and is very grateful to the members of the Fermilab 
Theoretical Physics Division and of the SLAC Theory Group for their 
kind hospitality during parts of these studies. This work was supported 
in part by the Department of Energy, Contract No.\ DE-AC02-76CH03000.


\begin{thebibliography}{99}
\bibitem{dighethank}We are indebted to Amol Dighe who invented this notation.
\end{thebibliography}
\end{document}